# The Extreme Physics Explorer and Large Area Micro-Channel Plate Optics (2011 SPIE V8147-55)

Michael Garcia*[a], Martin Elvis[a], Jon Chappell[a], Laura Brenneman[a], Daniel Patnaude[a], Ian Evans[a],
Ricardo Bruni[a], Suzanne Romaine[a], Eric Silver[a]
Richard Willingale[b], George Fraser[b]
T.J. Turner[c], and Brian Ramsey[d]

[a]Smithsonian Astrophysical Observatory, 60 Garden St., Cambridge, MA USA 02138;
[b]Department of Physics and Astronomy, University of Leicester, University Road,
Leicester UK LE1 7RH
[c] Department of Physics, University of Maryland Baltimore County,
1000 Hilltop Circle, Baltimore, MD USA 21250
[d] NASA MSFC, Huntsville, AL USA 35811

## ABSTRACT

The Extreme Physics Explorer (EPE) is a concept timing/spectroscopy mission that would use micro-channel plate optics (MCPO) to provide 4m$^2$ effective area focused to ~1 arc-min onto an X-ray calorimeter. We describe science drivers for such a mission, possible designs for the large area MCPO needed for EPE, and the challenges of the large area MCPO design.

## 1. INTRODUCTION

The Extreme Physics Explorer (EPE) is a large area, high spectral and timing resolution and polarimetry mission designed to make tests of fundamental physics in the areas of strong field gravity and Quantum Chromo-Dynamics (Elvis 2004[1]). These tests are enabled via observations of bright black holes and neutron stars as described in the Astro2010 Decadal Survey "New Worlds, New Horizons" and in the baseline International X-ray Observatory (IXO) observing plan as outlined in Garcia et al. 2010[2] The background within the arc-min PSF is equivalent to a sky flux of 3 x 10$^{-14}$ ergs/cm$^2$/s (2-10 keV) so good spectra are obtainable for all of the ~20,000 sources in the ROSAT BSC (Voges et al. 1999[3]). By using very lightweight MCPO and a single cooling string calorimeter, the mission mass could be as small as 450kg-650kg, well within the bounds of Explorer class missions.

## 2. SCIENCE

### 2.1 General Relativity

The observational consequences of strong gravity can be seen close to the black hole event horizon, where the extreme effects of General Relativity (GR) are evident in the form of gravitational redshift, light bending, and frame dragging. The X-ray band is the only spectral region which allows sensitive probes of the structure of the inner disk. EPE will allow us to observe spectral features from matter orbiting at these innermost radii (Figure 1, from Astro2010 White Paper: ``Spin and Relativistic Phenomena around Black Holes", Brenneman et al. 2009[4]). Observations of super-massive black holes (SMBHs) with XMM-Newton have revealed evidence of hot spots on the disk that light up in the Fe-K$\alpha$ fluorescence line, allowing us to measure their motions (Turner et al 2002[5]). These hot spots are expected from the most recent MHD modeling of disks (see Brenneman et al. 2009[4] and references therein), and have been seen in a few objects where the conditions are particularly favorable (e.g., NGC 3516, NGC 3783). Each hot spot is, in effect, a test particle and follows (to 1%) a Keplerian orbit around the black hole. The emission from these hot spots appears as arcs in the time-energy plane (Figure 1, bottom left). GR makes specific predictions for the form of these arcs, which are highly dependent on the mass and spin of the black hole and the inclination of the accretion disk. The time vs. velocity measurement of a single arc allows determination of these parameters. EPE observations of a single AGN will allow determination of the GR parameters from hundreds (Figure 1, bottom right) of hot spots in the disk over the course of a typical observation. Each measurement should yield the same mass, spin, and inclination therefore allowing sensitive measurement of the metric as predicted by GR under the strong field limit. As shown in Figure 2, there are several dozen AGN bright enough for EPE to measure these time resolved orbits near the event horizon. EPE will allow these measurements on SMBHs from $10^8$ solar masses down to (for the first time) $10^6$ solar masses, therefore covering the bulk of the range of masses typically associated SMBHs.

*garcia@cfa.harvard.edu; phone 1 617 495 7169; fax 1 617 495 7356; hea-www.harvard.edu/~garcia/

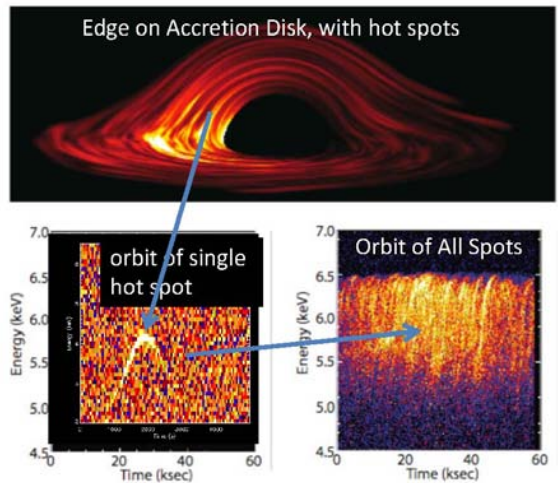

Figure 1: MHD Simulation of accretion disk around black hole, showing the relativistic Doppler shifts from hot spots orbiting near the event horizon. From Brenneman et al. (2009[4]).

As well as measuring the orbits of matter near the event horizon, EPE will allow measurements of photon orbits near the event horizon, a fundamentally different probe of GR in the strong gravity regime. This will be done via reverberation mapping; i.e., measuring the response of the Fe-K$\alpha$ fluorescence line flux to variations in the illuminating continuum. Fundamentally different and therefore independent probes of the structure of the space-time metric under strong gravity will come from measurements of quasi-periodic oscillations (QPOs) in black holes and from polarization measurements.

Time-averaged Fe-K$\alpha$ line shapes (as opposed to the time-resolved studies above) can give accurate measurements of black hole spin if at least 20,000 to 100,000 counts are collected in the X-ray spectrum. At $10^{-13}$ erg/cm$^2$/s (2-10keV) the EPE count rate will be ~1 c/s, so there are ~1000 AGN in which reasonable EPE exposures will yield the required number of counts and therefore will yield the black hole spin. However the interpretation of these line shapes is complicated by time-variable absorption from disk winds with a variety of ionization states and speeds. EPE can provide accurate measurements of absorption lines in these outflows, giving us the excellent constraints on total mass loss rates for the X-ray gas and also constraining the wind launch radii. Acceleration mechanisms could be ascertained from tracking line changes with time. If the wind is clumpy and has a rotational component of velocity, then absorption varies as clumps come across the line of sight. EPE will be able to track the motion of these clumps in the many of these AGN.

Assuming the effects of the absorbing wind can be understood, one will be able to determine spins from the line shapes in the entire sample of ~1000. If the effects are not resolvable reliable spins may be extracted from only 300-500 of the sample with weaker winds. These observations place strong constraints on SMBH merger and accretion models, as the spin is critically dependent on how SMBHs merge and accrete (Berti & Volonteri 2008[6]).

The baseline EPE design includes a polarimeter similar to that proposed for IXO (Bellazzini & Mureli 2010[7]). If the EPE MCPO were configured in a lobster geometry rather than Wolter-1, the cross-arm focus lends itself to being picked off via high graze angle reflectors which would have a natural polarization dependence and therefore provide an alternative path to polarization measurements. However, while the effective area at low energies (below 1.25keV) may be maintained in the lobster geometry, the area at high energies (>5keV) would be reduced by a factor of ~10 relative to a Wolter-1 optic of the same radius and focal length.

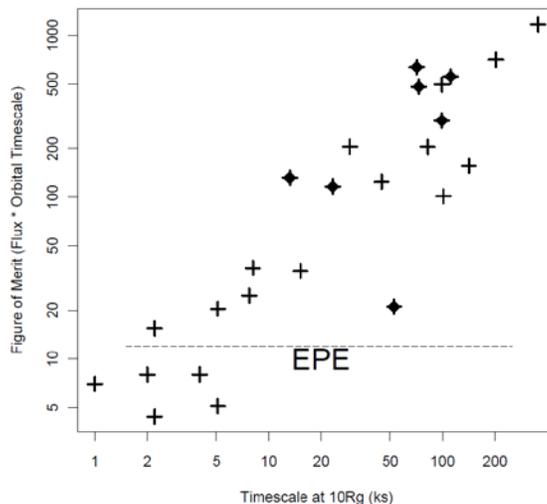

Figure 2: The orbital timescale at 10 Rg is plotted against the Figure of Merit for measuring orbits of test particles near the event horizon. The figure of merit is a scaled product of the flux in the Fe-K$\alpha$ line and the orbital timescale, divided by the square root of the continuum. EPE will reach down to a FOM of ~12, which will allow the measurements to probe down to SMBHs with mass of $10^6$ solar masses. Crosses represent SMBHs with an assumed Fe-K$\alpha$ equivalent width of 200 eV, while diamonds represent SMBHs with measured Fe-K$\alpha$ line strengths. We can be confident in this approach because hot spots have already been detected in a few of the most massive SMBHs (Turner et al. 2006[8]) showing the signature of test particles (hot spots) orbiting near the event horizon.

Polarimetry offers another way to probe the space-time metric near a black hole. Because of the increase in temperature with decreasing radius in the accretion disk, the polarization effects due to GR will be energy dependent. Measuring the size of this effect allows determination of the black hole spin (Stark & Connors 1977[9], Connors & Stark 1977[10], Connors, Piran & Stark 1980[11]). Moreover the polarization angle of the radiation determines the orientation of the accretion disk on the sky (Sunyaev & Titarchuk 1985[12]). Since the black hole mass is known from binary orbit analysis or from reverberation mapping for AGNs, all of the black hole parameters (except charge) are then determined.

### 2.2 Neutron Stars and Magnetars/QCD

Neutron stars consist of the most dense matter known, beyond the density of atomic nuclei. The form of this matter is beyond the current knowledge of particle physics, and is governed by the quantum-chromo-dynamics (QCD) description of the strong nuclear force. Neutron stars may consist of exotic quark matter, hyperons, or Bose condensates of pions or kaons. The way to distinguish between these possibilities and test our models of the strong force is to measure the equation of state, the mass/radius relation for neutron stars. EPE will be able to do this for a ~dozen bright neutron stars which are believed to cover a range of mass (see the Astro 2010 white paper by Paerels et al 2010[13], ``The Behavior of Matter Under Extreme Conditions'').

There are several approaches to this, but all require large collecting area and are greatly enhanced with the energy resolution enabled by a calorimeter. The most direct approach is to measure the gravitational redshift of absorption lines due to heavy elements in the atmosphere of X-ray bursting neutron stars. This was thought to have been done in EXO 0748-676 (Cottam et al 2002[14]) but recent measurements of the spin rate of this neutron star place this result in serious doubt. However, the discovery of an 11 Hz bursting pulsar in Terzan 5 (Strohmayer & Markwardt 2010[15]) shows that there are suitable slowly spinning neutron stars. This remains a virtually unexplored area as the high collecting area and energy resolution required to do this during a single burst is not yet available: The EXO 0748-676 results was extracted from the sum of 30 bursts, but EPE will allow measurement of lines within a single burst. There is a sizable region of parameter space in which even rapidly spinning neutron stars are expected to show narrow atmospheric lines (Ozel & Psaltis 2003[16]; Bhattacharyya, Miller, & Lamb 2006[17]).

Bursts on neutron stars originate on a small region of the neutron star surface, therefore the flux measured during the initial rise shows rapid pulsations. The shape of these pulsations yields the surface gravity of the neutron star and, along with the measured period, the mass and radius separately. EPE has sufficient time resolution that even for the most rapidly spinning neutron stars one can phase bin the data from even a single burst on a small fraction of the spin period and remove the Doppler broadening, which greatly enhances the detectability of these intrinsically narrow lines. The unique combination of large effective area and high time/energy resolution that is provided by EPE will allow this very direct and model independent measurement. Of course, determining the Doppler shift immediately yields the radius (modulo sin(i)), which when combined with redshift yields the mass.

The magnetars are a small subset of neutron star binaries containing highly magnetized neutron stars. In these systems the magnetic field is above the quantum critical limit $B_c = 2\pi\, m_e^2\, c^3/he = 4.4 \times 10^{13}$ Gauss, reaching values up to $10^{15}$ Gauss (Thompson & Duncan 1995[18], 1996[19]). At these field strengths interesting new effects are predicted (Erber 1966[20], see e.g. Miller 2001[21] for a recent review). For example, the transmission of light through a super-critical field is polarization dependent (e.g. Broderick & Blandford 2003[22]), and a characteristic polarization signal is expected. Another novel effect expected at $B_e$ is that the cyclotron radius should be smaller than the Bohr radius. As a result there is a squeezing of the wave packet along the field lines. This squeezing shifts the atomic energy levels, leading to hydrogen Lyman-α being shifted up into the soft X-ray regime (Ho et al., 2003[23]). The high effective area, high spectral resolution, and polarization capabilities of EPE will allow it to make the most sensitive measurements of these effects to date.

### 2.3 Flow-down of Key Science Requirements

**Area**: In order to time resolve the orbits of hot spots around SMBH we must collect sufficient photons within a single particle orbit. The target AGN have X-ray fluxes of $10^{-2}$ to $10^{-3}$ c/cm$^2$/s and orbital timescales of 1ks to ~100ks. Only with effective areas (>2m$^2$ at 6keV) can we collect sufficient numbers of counts to time resolve the orbits in a large sample (i.e., >20) of AGN and cover the full range of masses from $10^8$ to $10^6$ solar masses.

**Energy Resolution:** The higher the energy resolution the more accurately the orbits near the event horizon can be traced. It is expected that MHD effects in the disk will become apparent once orbital velocities are measured to <1%, so this sets a natural limit to the desired energy resolution. Time averaged spectra of AGN show complex absorption spectra which can complicate the interpretation of the relativistic iron line shape. Energy resolution of <1% is also required to understand these absorption spectra. The complex and likely variable intrinsic absorption in AGN complicates the measurement of the shape of the Fe lines and therefore determination of the spin. The spectral resolution of X-ray CCDs is typically insufficient to resolve all the absorption lines, but the <1% resolution of a calorimeter allows the lines to be resolved.

**Angular Resolution:** This is set by a combination of the plate scale of the optic and the need to avoid the confusion limit. At 40m focal length the plate scale is ~1 arc-min/ cm. The proposed IXO calorimeter is 3.1 cm on a side, so a 1 arc-min beam would nicely fill the calorimeter and still leave room for contemporaneous background measurements, be well above the confusion limit for typical EPE target fluxes, and allow for jitter in the mirror-detector system.

**Bandwidth:** The Fe-Kα line extends from ~2keV to 7keV, and the Compton hump extends from ~10keV to 40keV. In order to measure the continuum well it is necessary to have a band-pass that extends at least into this hump, so that the continuum above the Fe-Kα line can be determined. In order to observe the hydrogen Lyman-α series shifts expected in magnetars we need to extend the band to 0.5keV through 2.0keV, and in order to observe the gravitational redshifts of absorption lines due to heavy elements in bursting neutron stars we must cover the 0.5keV to 1.0keV band. Therefore the EPE bandwidth must be 0.5keV to at least 10keV, and extending up to 40keV would be desirable.

**Mass:** While not a 'scientific' requirement, low mass is a requirement for all space missions. Given the area requirement of $2m^2$ at 6keV the optic for EPE must go beyond current optics in terms of areal density, the ratio of effective area to mass. At the same time the angular resolution requirements for EPE are modest. Figure 3 compares MCPO to current X-ray optics and shows that it is the clear choice for this application.

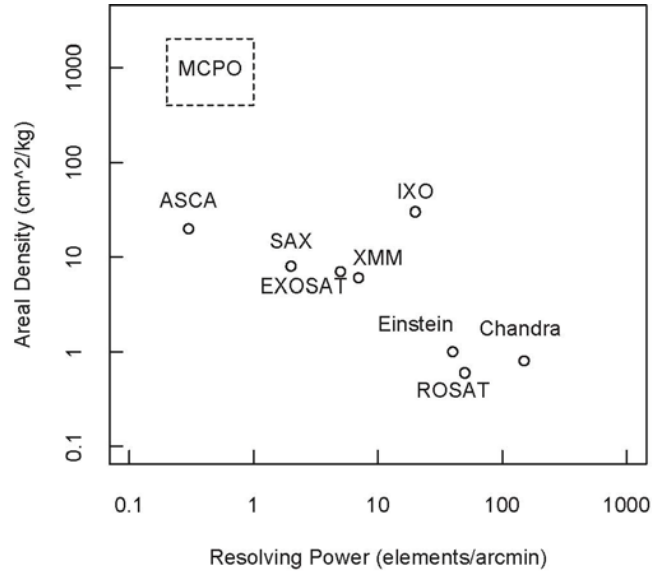

Figure 3: The resolving power vs. areal density for current X-ray optics, and the range allowed by MCPO. We will push towards the right side of the MCPO bounds, towards 1 arc-min resolving power. Adapted from Bavdaz et al (2010)[24].

**Summary:** Currently the only devices with the required energy resolution are X-ray gratings and micro-calorimeters. Gratings typically require angular resolution <<1 arc-min in order to yield <1% energy resolution, so are not compatible with the ~arc-min angular resolution envisioned here. Micro-calorimeters are well matched to the requirements. Clearly focusing optics are required in order to provide several $m^2$ of collecting area onto the micro-calorimeter, and the arc-min PSF for EPE is a plus as it distributes the counts over many calorimeter pixels and therefore allows the detector to count at high rates.

## 3. MISSION

EPE has a baseline 40m focal length and a heavy calorimeter at one end and light optic at the other, so would be subject to strong gravity gradient torques in low earth orbit. An L2 orbit is the most viable option as it would allow long observation lengths and would reduce these torques. In order to reduce the offset between the center of mass and center of (solar) pressure, the solar panels would likely be located near the calorimeter end of the spacecraft. The calorimeters from Astro-H and IXO have redundant cooling channels in order to reduce risk, but this also increases mass. Astro-H design mass is 366kg, IXO is 263kg (Bookbinder 2009[25]). Assuming a successful Astro-H mission it would be reasonable to consider a single cooling string system for EPE, which could reduce the mass to ~200kg. Given the range in area/mass covered by MCPO, we would expect the EPE optic including support structure to weight 25kg to 125kg, for a total payload mass of 225ks to 325kg. Spacecraft mass is typically about equal to the payload mass, so a total mass of 450kg to 650kg is expected. This is well within the MIDEX mass range, for example the mass of SWIFT is 1500kg, and

NUSTAR 300kg. The mass of the flagship class IXO is estimated at ~4400kg, not including margin (Bookbinder 2009[25]).

The 40m focal length will be enabled with a set of Able 'Adam' masts, which are available up to 60m in deployed length and have a deployed length accurate to 1.3mm (www.aec-able.com/Booms/adam.html).

Fairings with internal diameters of 4.6m are available for the Falcon 5 and Atlas V launchers, and may allow the 2.1m radius optic considered here to be accommodated directly. Alternatively, the optic could be deployed on orbit, in which case its very low mass will be a plus.

## 4. OPTICS

### 4.1 Motivation

Grazing incidence X-ray optics make large gains in efficiency as the substrate thickness is decreased, therefore allowing the aperture filling factor to increase. Ideally the projection of the reflecting surfaces onto the entrance aperture plane would completely fill that plane. The nested shells of the optics are typically separated by at least the thickness of the substrates. If the separation between shells (=D) is equal to the substrate thickness, the maximum fraction of the aperture which can be filled is 50%, so larger separations and correspondingly longer optics are desirable. The reflectivity of high-z coatings limits the graze angles $\phi$ of the incoming beam to less than few degrees, so the length L of the optic is typically $\sim D/\sin(\phi)$. For a 0.75 degree graze angle, equal to the point where the reflectivity of 300 angstroms of iridium drops to 50% for a 6.0keV X-ray, the sub-straight length L=76D. This length sets the thickness and ultimately the mass of the optic.

Successive generations of X-ray optics have moved steadily in the direction of increasingly full apertures (see Figure 4). For example, Chandra uses ~inch thick glass as a substrate for its iridium reflective coating, XMM uses 0.3mm to 0.7mm nickel substrates (enabling higher aperture filling) and the IXO segmented glass optic uses 0.4mm glass substrates and therefore fills an even higher fraction of the entrance aperture. The IXO silicon-pore optics (SPO) are thinner still at 0.17mm.

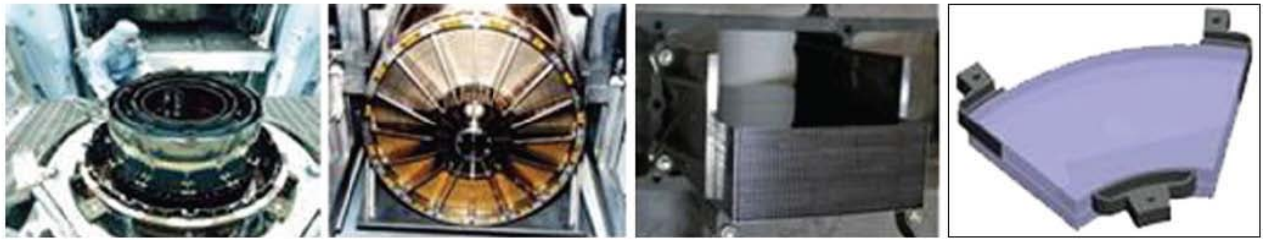

| Chandra | XMM-Newton | IXO-SPO | MCPO |
| --- | --- | --- | --- |
| 18500 kg/m$^2$ | 2300 kg/m$^2$ | 200 kg/m$^2$ | 5-25 kg/m$^2$ |
| ~250mm | 3mm-7mm | 0.17mm | 0.006mm |

Figure 4: Mass to effective area ratios, and sub-straight thickness for existing and planned X-ray optics. MCPO yield 5-25kg/m$^2$, an order of magnitude improvement over IXO. (From Bavdaz et al 2010[24])

The thinnest substrates currently available are those of the square pore micro-channel plate optics (MCPO). These are typically 6μ-2μ thick, so a factor of 30 to 100 times thinner than the current state-of-the art for the IXO optics. This suggests MCPO could be 30 to 100 times lighter than the current state-of-the art. The mounting structure may be as massive as the optic, and when included may put MCPO in the 5 to 25 kg/m$^2$ range. Because MCPO are stiff when compared to thin foil optics they may come in at the low end of this range. As the cost of a satellite mission scales closely with mass, this suggests that MCPO based X-ray missions would be considerably less expensive. Currently these optics are essentially made by hand in a very labor intensive process, which raises the cost above what this naïve scaling with mass would indicate. However, MCP are routinely mass produced at low cost for night vision use, which indicates that MCPO manufacture could similarly be scaled up and the costs decreased.

## 4.2 MCPO State of Art

While MCPO have long been envisioned for use in all-sky monitors in the Lobster-eye configuration (Fraser et al. 1992[26]), use of MCPO for arc-min focusing is in its infancy. Coating the very long, narrow channels of the MPCO with high-z materials in order to enhance reflectivity has just begun and the performance still uncertain; coating with multi-layers to go to high energies has not yet been tried. The most advanced focusing optic to date is MIX-T (Fraser et al. 2010[27]) on Bepi-Columbo, which has the requirement to achieve <9 arc-min HPD and uses MCPOs in a Wolter-1 conical approximation. The MIX-T optic has a 21cm entrance aperture and 1m focal length. Individual sections of the 36 wedge-shaped tiles for MIX-T have measured as good as 3 to 5 arc-min HPD. The individual MCPO sections for MIX-T are manufactured by Photonis, and then assembled onto the optic support frame at The University of Leicester.

## 4.3 The Path to 1 Arc-Min

Photonis is able to manufacture square channel plates with channel sizes ranging from 10μ-100μ and L/D ratios ranging from 30 to 500. The channels can be arranged in the traditional square lobster geometry or packed in a radial fashion (Figure 5). Plate thickness are typically 1mm to 10mm, and sizes range from 10mm to a maximum of ~10cm. X-ray HPD have been measured to be as good as ~15 arc-sec over a few individual pores (Mutz et al. 2007[28]), and ~20 arc-sec over a single multi-fiber (Collon et al. 2007[29]). However the process of fusing numerous multi-fibers into a large plate, and then slumping that plate to a spherical surface introduces irregularities in the alignment of the reflecting surfaces which limits the HPD to several arc-min (Collon et al. 2007[29], Price et al. 2002[30]). The fundamental limit to MCPO is on the arc-second level, as it is equal to the projected size of a single pore onto the focal plane.

Possible improvements to the manufacturing process have been identified (i.e., Collon et al. 2007[29]) and many of these may be possible to implement given the configurations needed for EPE. For example, one of the current limitations is the alignment of numerous multi-fiber bundles: for EPE larger 60μ pores are appropriate, and we may be able to work with smaller numbers of bundles (i.e., 3x3 or 10x10) rather than the 20x30 typical of the 20μ pore optics used in MIX-T. In addition the slumping process needed for the 1m focal length of MIX-T appears to increase the HPD - for EPE we may be able to work with individual 1cm square MCPO that are not slumped, but instead mounted on a curved frame. Fitting the square fibers into the wedge-shaped individual MIX-T optics can introduce its own packing error (Figure 5) therefore increasing HPD; we can avoid this for EPE by working with completely square individual optics.

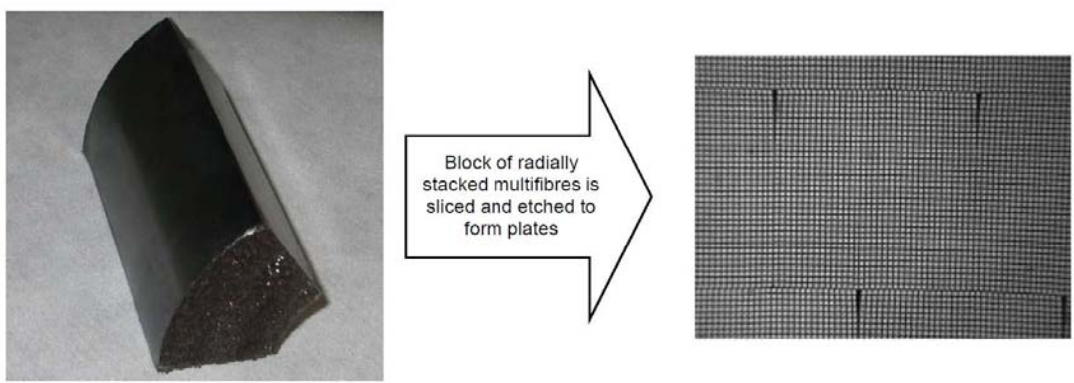

Figure 5: Radially packed multi-fibers may show deviations from perfect geometry at the boundaries of the multi-fibers, as shown in this microscope photo from Wallace et al. (2006)[31]. These deviations limit the PSF to >1 arc-min. Tiling square geometries at long focal lengths may mitigate these effects.

## 4.4 Basic Design Parameters of MCPO for EPE

EPE could be based on single or double (Wolter-1 approximation) reflection geometry and have a focal length from 20m to 40m. It could retain the traditional lobster geometry wherein all the optical surfaces are at 90 degrees to each other, or utilize square channel MCPO in a radially packed fashion (Figure 6) as in Bepi-Colombo. A 40m focal length may allow individual square MCPO of ~1cm to be arranged in radial tiles while still maintaining arc-min PSF, in an approximation of the true radial packing arrangement (Figure 6, right). This is because the projected size of an individual 1cm square MPCO is only 0.85 arc-min at the 40m focal length for EPE.

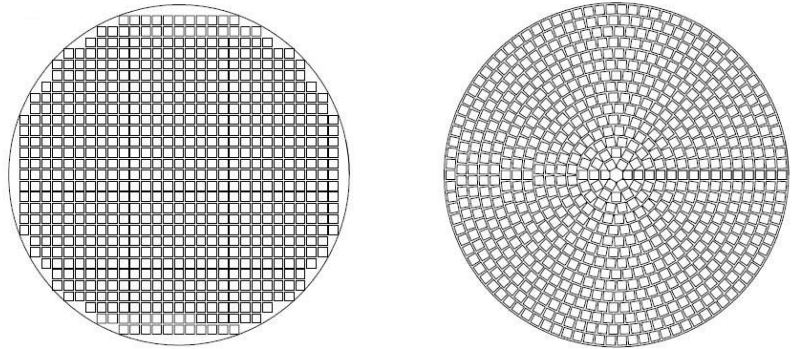

Figure 6: Traditional square-pack ``Lobster'' geometry (left) and radial-pack geometry as used in Bepi-Colombo (right). From Willingale et al. (1998).

The maximum radius of the optic is set by the need to limit the graze angle: for Ir reflectors, which are typically the optimum choice for providing good reflectance up to ~7keV; the reflectance falls below 50% for graze angles larger than 0.75 degrees. This sets the L/D ratio for the outer edge of the mirror to ~76. This L/D ratio is the same for a single (lobster geometry) or double bounce (Wolter-1) optic, but the outer radius of the mirror can be twice as large for the double bounce, as the rays are redirected at four times the graze angle rather than twice the graze angle. The larger mirror not surprisingly gives larger effective area. For the 40m focal length case, the outer radius is then 2.1m (two bounces) or 1.05m (single bounce). To retain complete aperture filling and minimize mass one would profile the optic by increasing L/D with 1/radius. The innermost L/D could go as high at 300, but could be limited to ~160 if required.

In calculating the effective area we assume the geometric area of the MCPO which intercepts the input beam, the reflectivity for 300 angstrom of Ir over the glass, and various geometric blocking factors, and we have assumed reflecting surfaces with 2 angstroms micro-roughness. While this is smoother than often found for MCPO reflecting surfaces (~15 angstroms, Prince et al. 2002[30]), it is consistent with the results obtained by magnetically assisted finishing of silicon optics (Riveros et al. 2011[39]). We include a 60% open area ratio, which is what the 20µ pore, 6µ webbing, MCPO for MIX-T have, even though with 60µ pores and the same webbing the open area ratio would improve to 83%. For Wolter-1 like designs we assume the second MCPO will on average not line up with the first, causing another geometric blocking factor of 20%. The structural model for the IXO slumped glass mirrors includes spiders as a mechanical support for the mirror shells, which adds an additional 11% to 24% blockage depending upon the shell position (Paul Reid, PC). We therefore assume an additional 25% blockage due to mechanical support because the structure needed for the very thin MCPO is likely to be higher than that needed for the IXO. Throughput at low energies is then 28% of the total geometric area for the Wolter-1 design (Figure 7). This seems reasonable when compared to the 39% for the IXO slumped glass optic design. However it may be optimistic given the 15% throughput estimated for the MIX-T design (Fraser et al 2010[27]). The open area ratio considered above has a strong effect and if applied to MIX-T would boost the predicted throughput of MIX-T to 25%, lending support for our 28% estimate for EPE.

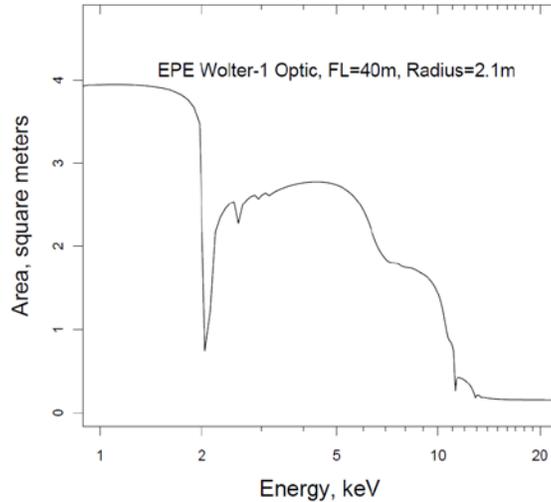
Figure 7: Effective area of EPE MCPO as described in the text.

## 5. DETECTOR BACKGROUND

The instrumental background will come from both cosmic (local and extra-galactic) and particle induced sources. We consider both below.

**5.1 Particle Induced**:

The instrumental background will come from both cosmic rays and solar wind particles. Unfortunately the background induced by these components in an X-ray calorimeter at L2 is rather uncertain. The only calorimeter yet flown in orbit, the Suzaku XRS, was in low Earth orbit and therefore partially shielded from both sources. However the elliptical orbits of Chandra and XMM brings them outside the shielding of Earth's magnetosphere, so the rates in the X-ray CCDs on board may yield some insight into rates at L2. As calorimeters and CCDs likely respond differently to particle induced background, a good proxy for the EPE calorimeter rates might be to scale the Suzaku rates by the ratio of the rates in the Suzaku CCDs to those seen in the Chandra and XMM CCDs.

In doing this, we follow the work of Smith et al. (2010)[32], who finds that the background in the X-ray CCDs at higher orbit is ~4x that seen in low Earth orbit. We therefore assume that the rate for the EPE calorimeter will be 4x that seen in the Suzaku XRS. The figure below shows the residual background in the XRS during a 37ks window between SAA crossings. Events due to particle crossings, such as those triggering the anti-coincidence shields and those which triggered multiple pixels at the same time, have been removed. Remaining events are likely due to secondary particles. These data were taken with the vacuum gate valve above the detector closed, so clearly cannot contain any cosmic X-ray background component. On the other hand, in actual operation this gate valve would be open, therefore changing the operating environment of the detector and possibly also the residual non-rejected, non-X-ray, background. The XRS background can be adequately described as a step function changing at 6keV, with $B(<6keV) = 6 \times 10^{-5}$ c/s/keV/mm$^2$ and $B(>6keV) = 2.5 \times 10^{-5}$ c/s/keV/mm$^2$ (Figure 8). The detector plate scale at 40m focal length is 1.17cm/arc-min, so within the 1 arc-min HPD PSF we would see $B(0.5-6keV) = 0.16$ c/s and $B(6-12keV)=0.08$ c/s, for a total of $B = 0.24$ c/s/PSF.

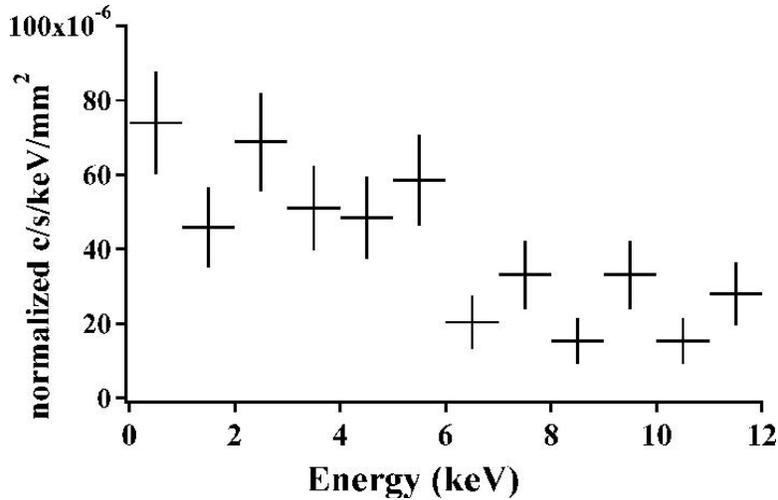

Figure 8: Residual background due to particles as seen by the Suzaku XRS. The residual background in the EPE Calorimeter is assumed to be ~4x this rate, due to the fact That the rates in the Chandra and XMM are ~4x that in The Suzaku CCD cameras. From Kelley et al. (2007)[33].

### 5.1 Cosmic X-ray Background:

The cosmic X-ray background (CXB) is focused through the telescope, consists of local (Earth Magnetosphere, Local Hot Bubble, and Galactic) components which are diffuse and largely dominated by emission lines with little continuum, and the extra-galactic background made up of unresolved sources, largely QSO and distant galaxies.

The solar wind interacts with Earth's magnetosphere, producing solar-wind charge exchange emission. This emission varies widely in time due to variations in the solar wind, and in space as the magnetosphere changes shape due to the pressure of the solar wind. An orbit at L2 (Figure 9) may well pass in and out of the magnetosheath, adding to the variability of this background component. Fortunately this component is dominated by narrow emission lines, so the high resolution of the EPE calorimeter will allow it to be removed. The same is true of the Local Hot Bubble emission, which is typically modeled as a ~$10^6$K plasma. The lines of O VII and O VIII at 0.57keV and 0.65keV are clearly detected in both components (Kuntz and Snowden 2004[34], Smith 2007[35]). The Galactic emission is strongest at low Galactic latitudes (<5 degrees), but again can be modeled as a soft thermal plasma and is line dominated, so can be removed.

The spectrum of the extra-galactic background is primarily a continuum due to AGN and therefore cannot be removed by ignoring narrow spectra ranges (as above). This CXB has been measured by De Luca and Molendi (2004)[36] over the 2-10keV range to have a power law shape with a photon index $\Gamma=1.41$ and a flux of $2.2 \times 10^{-11}$ erg/cm$^2$/s/deg. This would produce 0.019 c/s/arc-min square over 2-10 keV through the EPE Wolter-1 design mirror, about 2.5x that predicted for the IXO slumped glass mirror (Smith et al. 2010[32]). The rate from 1-2keV is depends somewhat upon Galactic absorption along the line of sight, but in an area of low absorption would be 0.012 c/s/arc-min square. So we find that the non-rejected particle background will dominate and is ~8x more than the diffuse CXB.

The confusion limit is the flux at which there is a significant (typically >10%) chance that one of the individual sources making up the CXB will be in the (arc-min) PSF. Given the logN/logS curve of Moretti et al. (2003)[37] there is a 10% chance of a source with a flux of $10^{-14}$ ergs/cm$^2$/s with and arc-min beam, or a 3% chance of a source with a flux of $3\times10^{-14}$ ergs/cm$^2$/s. This latter flux would produce a count rate approximately equal to the non-rejected particle background. A typical EPE observation therefore would not be confusion limited, but observations attempting to reach well below the background rate (for example, to $10^{-15}$ ergs/cm$^2$/s) would run into the confusion limit due to the arc-min PSF.

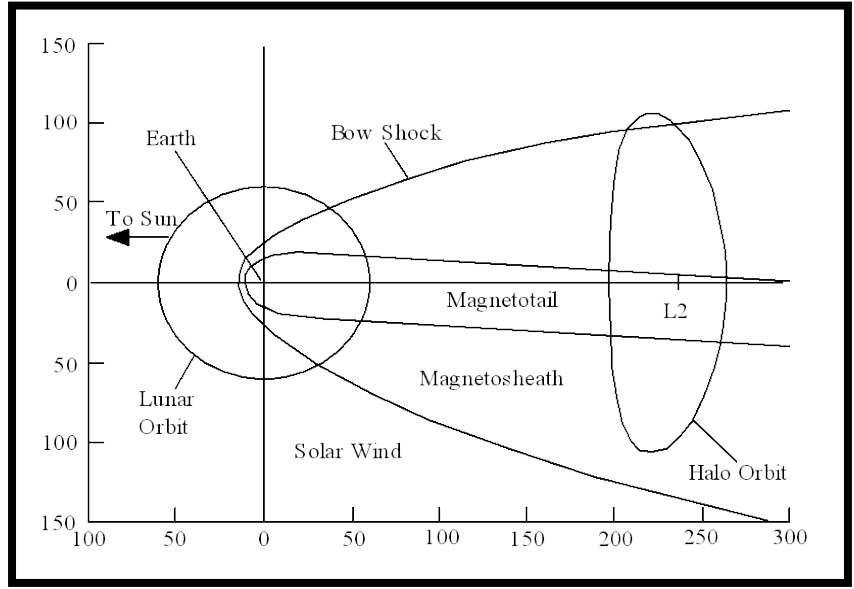

Figure 9: Cartoon of likely EPE orbit at L2, and geometry of magnetosphere. The orbits of both Chandra and XMM-Newtown are contained within the lunar orbit, but also pass through the magnetotail. EPE will likely experience a range of background due to solar wind charge exchange due to this geometry. From Sanwal (2008)[38].